\documentstyle[12pt,aj_pt4]{article}

\def\etal{{\it et al.\,}}
\def\logg{log {\it g}}
\def\teff{T$_{\rm eff}$}

\lefthead{Fulbright \etal}
\righthead{Abundances \& Kinematics I}

\begin{document}{
\title{Abundances and Kinematics of Field Halo and Disk Stars I:\\ Observational Data and Abundance Analysis}

\author{Jon P. Fulbright\altaffilmark{1}$^,$\altaffilmark{2}} 

\affil{UCO/Lick Observatory, Dept of Astronomy, University of California,
    Santa Cruz, CA 95064}

\slugcomment{Accepted for Publication in the September 2000 AJ}

\altaffiltext{1}{Present address:  Dominion Astrophysical Observatory, Herzberg Institute of Astrophysics, National Research Council, 5071 West Saanich Road, Victoria, BC V8X 4M6, Canada}
\altaffiltext{2}{Email:  Jon.Fulbright@nrc.ca}

\begin{abstract}
	We describe observations and abundance 
analysis of a high-resolution, high-S/N survey of 168 stars, most of which are 
metal-poor dwarfs.  We follow a self-consistent LTE analysis technique to 
determine the stellar parameters and abundances, and estimate the effects 
of random and systematic uncertainties on the resulting abundances. 
Element-to-iron ratios are derived for key alpha, odd, Fe-peak, r- and 
s-process elements.
Effects of Non-LTE on the analysis of Fe I lines are shown to be very small
on the average. Spectroscopically determined surface gravities are derived
that are quite close to those obtained from Hipparcos parallaxes.

\end{abstract}


\section{Introduction}

	The traditional explanation for the chemical evolution of the 
Galactic halo was put forth by \cite{t97} and is based on the 
differing products of the two main types of supernovae.  
Type Ia supernovae produce mainly Fe-group elements, while Type II supernovae
produce lighter elements (including the so-called `alpha' elements)
as well as some Fe-group and heavier elements.  Since the time between 
star formation and explosion differs between the two
types (Type II need $\sim 10^7$ years, while Type Ia need $> 10^9$ years), 
there is a time in which the enrichment is from Type II SN.
The stars created out of the ashes of
these early Type II supernovae will be relatively rich in the alpha (and
other light) elements until enough Type Ia supernovae can
explode to `dilute' the light elements with Fe-group elements.  
This overall pattern is seen observations of halo stars and clusters 
(see \cite{w89} and \cite{m97} for reviews), and
indicates that element ratios can be used as an indicator of the history
of a stellar population.

	Recent developments have shown that the chemical evolution of the 
halo is more complicated than the original \cite{t97} model.  \cite{ns97}
studied stars of intermediate metallicity ($-1.3 < $[Fe/H]$ < -0.5$)
and found that there was a number of stars on halo-like orbits that 
exhibited significantly lower [$\alpha$/Fe] ratios than the disk stars at 
similar
metallicities.  \cite{k97} found solar-like [$\alpha$/Fe] ratios in the 
metal-poor proper motion pair HD 134439/40, and \cite{c97} found
sub-solar [$\alpha$/Fe] ratios in the even more metal-poor star BD +80 245.

	From a suggestion in \cite{ns97},
\cite{h98} found that in a sample metal-poor 
halo giants, about a third of the stars on retrograde orbits showed low
values of [Na/Fe], compared to none of the stars on prograde orbits.
\cite{s99} followed a suggestion by \cite{c97} to study stars on orbits 
with large apogalactic radii, but concluded that the abundance ratios of
such stars were not significantly different than what is found  
in the rest of the halo population.

	The above studies do  indicate that some halo stars deviate from the 
traditional halo chemical evolution model. However, in all of the above works
only a limited number of stars were observed, so it is not readily possible to 
understand the true distribution of element ratios in the halo.  Similarly, 
systematic differences between abundance studies, due to differences in data 
quality, wavelength coverage, temperature scales, atomic data, 
model atmospheres, etc., make it difficult to combine the results from 
previous surveys to improve the situation.

	With this in mind, we have conducted a self-consistent survey
of metal-poor stars with the goals of determining the distribution
of element ratios within the halo, finding potential relationships between
stellar kinematics and these element ratios, and how these relationships
relate to the early chemical evolution of the Galaxy.  

        In this paper, we analyse the elemental abundance ratios
of a large sample of metal-poor stars using high-resolution, high-S/N
spectra.  In Section 2 we discuss the selection of target stars and the
observations.  Section 3 includes the process for selecting the absorption
lines and line data as well as the measurement of
the line strengths.  In Section 4 we use the line data to determine the
stellar parameters for the target stars and then compare the results to 
literature determinations.  In Section 5, we use these parameters 
and the line strengths to determine the abundances of the elements and 
estimate the uncertainties of the resulting abundances, while in Section
6 we discuss the possibility of non-LTE effects on the analysis.

        In following papers in this series will focus on the scientific 
interpretation of the results of this analysis.  Paper II will describe the 
trends of the elements with respect to [Fe/H] and to each other, as well as 
exploring relationships between the stellar kinematics and 
abundances.  Paper III will use the derived metallicities of nearby stars 
to examine the subdwarf distance scale.  Descriptions of further research with 
this data set will be included in these papers.

\section{Star Selection, Observations, and Reductions}

\subsection{Target List}

	Our target list was created from multiple literature sources, 
selecting for metallicity, membership in the Hipparcos catalog, and 
observational considerations. 
However, this list is subject to selection biases common between the source
surveys.  In particular, many of the source papers select stars by
kinematical criteria, potentially ignoring metal-poor stars with small
proper motions.  Although using multiple source surveys with different
selection techniques may help eliminate this problem, we make no claims
that this work is without selection effects.

        Several sources contain lists of [Fe/H] values determined by
spectroscopic means and include:
\cite{cay97}, \cite{clla}, \cite{b80},
\cite{p93}, \cite{bsl}, \cite{f95},
and \cite{cv97}.  Other sources used here, based on multicolor photometry 
 are \cite{sf87},
\cite{sn88}, \cite{s93} and \cite{o94}.  
\cite{sf87} observed stars in UBV filters, and we used their calibration 
of the $\delta$(U-B) index to estimate [Fe/H].  The other sources use 
{\it ubvy} photometry, and we use the calibration of \cite{sn89} 
to convert to [Fe/H].  For all of the above sources, stars with 
reported [Fe/H] $\lesssim -0.8$ 
were placed on the preliminary target list.  We also included a list of 
kinematically-selected stars from the Hipparcos catalog \cite{hip} from
Hanson  (private communication) and several stars suggested by  
N. Reid (private communication) and R. Peterson (private communication).
The final target list includes over 400 stars.

\subsection{Observations and Data Reduction}

        The stellar spectra were obtained between August 1994 and May 1999.
Table 1 gives the details of each run, and Table 2 lists the basic
observational data on each star.  Generally, the observations were taken
at spectral resolution R $\sim$ 50000, with a minimum S/N of 100 (per pixel).  
The S/N level at 5500 \AA$\;$ is listed in Table 2.  

        The majority of the data was obtained using the Shane 3-m
telescope and the Hamilton spectrograph at Lick Observatory.  
In 1994 and 1995 the observations were obtained by M. Bolte, K.
Wu, and M. Shetrone.  The 1994 runs used a 800x800 pixel CCD, which 
limited the wavelength coverage.  Since that time a 2048x2048 CCD 
was used allowing full wavelength coverage.  Data from the 
ESO 3.6-m and CASPEC spectrograph was kindly provided by M. Shetrone, who 
observed the stars during engineering time.  The stars observed with the 
10-m Keck 1 telescope and HIRES spectrograph (\cite{v94}) in 1998 
and 1999 were obtained during twilight, poor weather (in 1998) or between 
observation fields (in 1999).  

        Data reduction was done with the use of programs in the IRAF 
package `echelle'.  The methods used are the same as
those used in \cite{f99}.  Only minor adjustments to the
procedure were required to reduce the data obtained using spectrographs other 
than the Hamilton.

\section{Line List and EW Measurements}

\subsection{Line Selection and gf-Values}

	A key problem in the analysis of stellar abundances is the need for
quality atomic data.   As the overall goal is to have the best 
possible abundance determinations, our guiding philosophy for line
selection is to identify a set of self-consistent absorption lines that
have high-quality relative gf-values and are easily measurable in the 
spectra of stars observed here (mainly metal-poor dwarfs).

	The final line list is given in Table 3.  Some of the gf-values have
been altered from their original published values.  The Fe I gf-values from
the Oxford group have been increased by 0.04 dex, following a 
suggestion by \cite{mcw95b}.  The \cite{ox80b} Fe II gf-values were 
calculated using a solar $\log{\epsilon{\rm (Fe)}} = 7.67$\footnote{In this 
paper, we define $\log{\epsilon{\rm (X)}} = \log{\rm n(X)} + 12.0$.}.  
We therefore raised the gf-values by 0.15 dex to reflect our adopted solar 
Fe abundance,  $\log{\epsilon{\rm (Fe)}} = 7.52$.  
The \cite{m83} Fe II gf-values have
been shifted by $+0.11$ dex to agree with the \cite{ox80b} scale.
Poorly-performing lines from the references in Table 3
were eliminated through an empirical test involving 16 high-S/N spectra
of well-studied stars.  Consistently deviant, strong and/or 
blended lines were removed through an iterative process.  

        As an independent test, we measured the EWs of the line list in the
two solar spectra taken in this survey.  We then used the traditional  
solar parameters\footnote{We use [Fe/H]$_{atm}$ to designate the overall
abundance of the input atmospheric model.  As will be discussed later, this 
may not be the same as the derived iron abundance ratio of the star, [Fe/H].} 
(\teff{} = 5770 K, \logg{} = 4.44, [Fe/H]$_{atm} = 0$ and $\xi$ = 0.84 km/s) 
to determine solar abundances.  
The results are exhibited in Table 4.  As can be seen,
we recreate the solar abundance ratios fairly well.  The results of the 
analysis also improves between the lower-quality October 1998 spectra 
(S/N $\sim 80$) and higher-quality May 1999 spectra (S/N $\sim 220$).  
It should also be noted that most of the lines analyzed here are their 
strongest in the Sun, so the problems of damping and unknown blends should 
be the largest in the Sun.

	Hyperfine splitting (hfs) effects were taken into account for the 
Li I and Na I D lines and the lines of Ba II and Eu II.  We assumed solar
system isotopic ratios for these elements.  Other odd-z elemental lines were
treated as single lines in this analysis.

	Except for Ba, the \cite{u55} approximation 
was used to calculate the van der Waals damping constant.
It was found that multiplying the
damping constant by some factor generally decreased the \teff{} obtained by
the Fe I excitation potential plot significantly.  As the weak lines used
for the analysis generally did not show deviations from Gaussian profiles
(see below), it was felt that these lines should not be affected
significantly by damping.  For Ba, however, it was found that it was necessary 
to multiply the Unsold approximation by 5.0, as suggested by \cite{gs94}.

        To test this assumption, we rederived the stellar parameters for 
eight dwarfs after eliminating all Fe lines stronger than 50 m\AA.  For the 
most metal-poor stars, there was little, if any, effect.  This was partially 
due to the small number of lines stronger than 50 m\AA$\;$ measured in 
these stars.  As the metallicity increased, there was a trend indicating a 
preference for slightly lower \teff{} and \logg{} values.  For dwarfs 
with [Fe/H] $> -1.0$, the effect was to prefer \teff{} values $\sim$75 K 
cooler and surface gravities $\sim$0.1 dex lower.  This trend was smaller 
for stars with lower surface gravities.  There was very little effect on 
[Fe/H]$_{atm}$ and no consistent effect on $\xi$.

\subsection{EW Measurements}

        Equivalent widths were measured using the 
`splot' program in the IRAF `echelle' package.  
Generally Gaussian fitting was
used to measure lines, as testing with ThAr lamp spectra showed that
weak lines were well fit by Gaussian profiles.  Direct integration was
used for the stronger lines (EW $\gtrsim 100$ m\AA).
Direct integration was also used when a line was
known to be one affected by hfs.

        Lines that were blended with telluric features, night sky emission
lines or the wings of strong stellar lines were not measured, nor were very 
strong lines and lines near or on bad pixel regions.
For most lines measured on most spectra with average S/N,
those measurements with EWs $\gtrsim 10$ m\AA$\;$ are thought to be
reliable, although this limit varies depending on S/N, line placement 
with respect to the blaze, and local spectral contaminants.

        In total, 41256 EWs for the 17 elemental species on 191
spectra were measured.  For stars with multiple observations, a single EW
list was created for the star by combining the EWs from each observation and
was used in the determination of stellar parameters and abundances.  Note 
that the EW measures for the individual observations were used in creating
the line list (section 3.1), the EW comparisons (section 3.3) and the 
determination of uncertainties (section 5).
These equivalent widths are not included
in this publication, but are available electronically from the Astronomical
Data Center (ADC).
 
\subsection{EW Comparisons}

        Figure 1 compares EW measurements between
seven stars observed at both Lick and McDonald.
The stars were observed 18 times (11
times at Lick, 7 times at McDonald), yielding 443 common line
measurements (counting multiple Lick observations independently).  
The average offset,
$\langle EW_{Ham} - EW_{McD} \rangle = -0.6 \pm 0.2$ m\AA$\;$ (sdom) is small.

        The observations from ESO and Keck do not have any stars in common
with any other observation run, so no such comparison is possible.
However, many observations have been made of common stars between
Keck and Lick, namely in the work of the Lick/McDonald group (\cite{matt96},
for example) and \cite{jj99}.  The
comparisons made in these papers show that there is not a systematic
difference in EWs obtained with these spectrographs.  No known study linking 
the ESO CASPEC spectrograph with either the Hamilton, HIRES or Sandiford 
spectrographs could be found.  

        Figures 2(a)--(d) compare EW measurements
for stars in common with other studies.  In all four cases,
the average offset is less than 1 m\AA.  In Figure 2(c) there are are about 
a dozen EWs for which the EWs from \cite{ns97} are noticeably 
greater than those measured from this study.  These EWs all come from 
the spectra of HIP 59750 (= HD 106516) taken at Lick in April 1999.  The 
other EWs in common for this star between the the \cite{ns97} list 
match well.  It is not believed that the different chip used in the April 
1999 run is the cause of this effect.  There were four stars 
in common between the April 1999 Lick run and the January 1999 McDonald run.
If there was a sizable offset between these runs, it would show up in
Figure 1, but only a few slightly weaker lines from HIP 59750 can
be seen in the 40--80 m\AA$\;$ range.  These anomalous measurements were 
discarded when creating the final EW list for this star.

\section{Stellar Parameters}

\subsection{Deriving the Parameters}

	The basic stellar parameters (\teff{}, \logg, [Fe/H]$_{atm}$, and 
$\xi$) were determined using the Fe lines in the spectra.  Before analysis,
Fe lines stronger than $\log(EW/\lambda) = -4.80$
($\sim 75$ m\AA$\;$ at 5000 \AA) were eliminated from both the Fe I 
and Fe II lists.  This limits the effects of damping on both line measurement 
and abundance analysis.  There were strong-lined stars, however, in which it 
was necessary to use stronger lines  in order to determine the parameters.  
Even in these cases the strongest Fe lines used were limited to 
$\sim 100$ m\AA. 

        We use the LTE abundance program MOOG (\cite{s73}) to
derive all of the abundances and synthetic spectra in this study.  We use 
Kurucz (http://cfaku5.harvard.edu) model atmospheres.
These atmospheres were computed
using solar abundance ratios and convective overshoot.  For atmospheres 
between grid points, a program was used to interpolate the values of
$\rho x$, temperature, gas pressure, and electron density for each layer 
within the atmosphere.  

	Since the assumption of solar abundance ratios does not hold in 
metal-poor stars, the value of [Fe/H]$_{atm}$ was set slightly higher than
the measured [Fe/H] value for the star.
This procedure is designed to simulate the increased supply of 
electrons ionized from the usual excess of alpha elements.  These extra 
electrons contribute significantly to the
${\rm H}^{-}$ opacity in the atmospheres of these stars.  

        We set the stellar parameters using an iterative process. 
First, the microturbulent velocity ($\xi$) is adjusted 
so the iron abundance given by the strong Fe I lines is the same given by
the weak lines.  Then the value of \teff{} is adjusted so the Fe I lines with 
high excitation potential (EP) give the same iron abundance as those with lower
EPs. Finally, the value of \logg{} is adjusted so the iron abundance given by 
the Fe I lines matches the iron abundance given by the Fe II lines 
(within $\sim 0.03$ dex in most cases).  This process is
iterative, with small changes being made to the parameters between each
step.

        As this is an iterative process, it is necessary to have a first
estimate of the parameters.  For \teff{}, we used the color-based \teff{} 
scales of \cite{c83}.  We gave preference to the $(V-K)$
relationship when the photometry was available.  Otherwise we used
the $(b-y)$ relationship from \cite{c83} or the $(B-V)$ tables given 
in Appendix B of \cite{g92}.

     We chose the initial value of \logg{} using the photometric \teff{} and 
the star's Hipparcos parallax.  For the stars here,
the mass was taken to be $0.8 M_{\odot}$, and the bolometric corrections were
taken from \cite{a95} for the dwarfs and Worthy (private communication)
for the giants.  The uncertainty in \logg{} from this method is dominated by 
the uncertainty in $\pi$, which means 
this method is only suitable for nearby stars with reliable parallaxes 
($\sigma_\pi/\pi <  0.2$), i.e. dwarfs.

        We set the initial value of [Fe/H]$_{atm}$ to slightly higher than 
the values given by the source papers, and the initial $\xi$ was set 
to 1.2 km/s.  The initial guesses of \teff{} and \logg{} plus the final 
adopted atmospheric parameters are listed in Table 5.

        This procedure was adjusted for $\sim 20$ giants in this survey. 
Strong lines and molecular features in several regions made it difficult to 
create a full set of reliable Fe I lines.  In these cases, we exploited
the known sensitivity of [V/Fe] to changes in \teff{}.  
This is similar to the method described in \cite{i99}, where the 
authors find a change in a giant's \teff{} by 125 K would result in a change 
in the [V/Fe] of $\sim 0.3$ dex.  As an Fe-peak element, [V/Fe] is expected to 
be close to solar over a wide range of metallicities.  This value has been 
observed in globular cluster giants (see Ivans \etal), field giants
(Johnson 1999) and dwarfs from this study.

        In these giants, we first attempted to use the Fe lines to set the 
stellar parameters as above.  When this was completed, if [V/Fe] $\sim 0$ 
(as was the case for stars with \logg{} $\gtrsim 2.0$), no further adjustments 
were made.  Otherwise, we changed the value of \teff{} as little as necessary 
to obtain [V/Fe] $\sim 0$ (usually upwards by $\sim$ 75--125 K).  The resulting 
compromise parameters usually left a slightly larger disagreement between 
Fe I and Fe II than was seen in dwarfs.

        In the few giant stars in which no V lines could be measured, we used
[Ni/Fe] and [Cr/Fe] in a similar way. 
However, both of these abundance ratios are less sensitive to \teff{}
than [V/Fe], and the value of the [Cr/Fe] ratio is known to decrease in 
very metal-poor stars
(McWilliam 1997).  Therefore [Cr/Fe] was not used for giants with 
[Fe/H] $< -1.5$.  Using these elements as backups was also helpful 
in the case of a
few stars with strong V lines, as the effects of hyperfine splitting on V
were not accounted for in this study.

        Also included in the data analysis were the stars of \cite{s99}, as
these high-velocity stars increase the number
of stars on extreme orbits in this survey.  The Keck-HIRES EWs from that 
paper were used with the line list in this study. 
For these stars the initial estimates for the stellar parameters 
were those given by \cite{s99}. 

\subsection{Comparison of Stellar Parameters to Literature Values}

        As a check of the parameters used here, we compare our values to
those determined by other studies.  \cite{a95} used the IR flux  
method (IRFM; \cite{bs77}) to determine the \teff{} values for a number of 
stars.  Figure 3(a) compares the 55 \teff{} determinations common between 
the two studies.  The average offset is 
$\langle T_{eff} - T_{Alonso} \rangle = -38 \pm 20$ (sdom).
There does not seem to be a slope over the $\sim 2000$ K range for which 
common \teff{} values are available.

       \cite{clla} also determined \teff{} values for a 
large number of stars using photometric indexes.  Figure 3(b) compares 
the 68 determinations in common.  The average offset is 
$\langle T_{eff} - T_{CLLA} \rangle = -12 \pm 14$ K (sdom).  

       Figure 3(c) shows the comparisons between the \teff{} values determined
by $V-K$ and $b-y$ colors using the \cite{c83} relationships.  The average
offset for the $V-K$ colors is 
$\langle T_{eff} - T_{V-K} \rangle = +36 \pm 15$ K (sdom, N = 105), and for 
$b-y$ it is $\langle T_{eff} - T_{b-y} \rangle = +109 \pm 14$ K (sdom,
N = 140).  It should be noted that no reddening corrections were made before 
calculating \teff{}, and the deviant points at \teff{} $\lesssim 4500$ K are 
giants.  For example, at $b-y$ = 0.50 mag, an E$(b - y) = 0.05$ mag 
will decrease the photometric \teff{} by 250 K.  Another possible 
explanation may be that these relationships were derived to fit dwarfs, 
and differences in the atmospheres of giants (e.g., stronger lines for a 
given \teff{}) could lead to changes in the \teff{}-color relationship.

        Figure 3(d) compares the \teff{} values determined here against the
published \teff{} values from several studies.  Overall, the other studies' 
\teff{} values seem to be $\sim$ 50--100 K warmer.

        Figures 4(a) and (b) compare the results of \cite{g96}
against the parameter values derived in this study.  \cite{g96}
re-analyzed literature data using \teff{} values derived from photometric 
colors and \logg{} values derived from ionization equilibrium.  The mean offset 
in \teff{} (this study $-$ theirs) is $-90 \pm 19$ K (sdom, N = 66), while 
the mean offset in \logg{} is $-0.22 \pm 0.05$ dex (sdom, N = 66). 
As seen with the comparisons to the \cite{c83} temperatures, the \teff{}  
values we derive for cool giants (\teff{} $\lesssim 4500$ K) 
are warmer than what was cited by \cite{g96}.
This is not seen when our \teff{} values for giants are compared with 
those derived by others using similar spectroscopic techniques,
(see Figure 3d), but is seen in the comparison to \cite{c83} photometric
\teff{} values.
Therefore, we believe there is a systematic difference between 
the photometric and spectroscopic \teff{} scales when applied to cool giants.  
These giants are also the ones in which the value of \teff{} was adjusted to
make [V/Fe] $\sim 0$, but the magnitude of the changes is smaller than 
the discrepancies seen.

        The final comparison made here is between the value of [Fe/H]
determined in this study and the [m/H] values presented in the \cite{clla} 
survey (Figure 5).  The latter were determined by chi-square fits of
high-resolution, very low-S/N spectra to a grid of
synthetic spectra.  The average offset for the 83 points is 
$\langle$[Fe/H] $-$ [m/H]$\rangle = +0.03 \pm 0.02$ dex (sdom, N = 71), 
which is surprisingly good agreement considering the different techniques
employed.  

\section{Abundances and Error Analysis}

        With all of the necessary ingredients in place, it is now possible
to calculate the abundances of the remaining elements.  This was
generally straightforward.  The line data were analyzed by MOOG using the
adopted atmosphere for that star.  The MOOG routine `blends' was used
for analyzing lines affected by hyperfine splitting.  For each star
the final abundance for a given species was the straight mean of the individual
lines.

        To compute the ratios of the elements with respect to the
Sun, we adopt the solar abundances of \cite{ag89}.  The sole
exception is Fe, for which we adopt $\log{\epsilon{\rm (Fe)}} = 7.52$.  
The derived abundance
ratios are given in Table 6.  The [Fe/H] given is derived from the mean
of Fe I and Fe II.  The [Ti/Fe] given is also derived from the mean of
Ti I and Ti II, except that the value of $\log{\epsilon{\rm (Ti I)}}$ 
was increased by 0.11 dex.  This offset was derived from the average difference 
between $\log{\epsilon{\rm (Ti I)}}$ and $\log{\epsilon{\rm (Ti II)}}$ for 
several stars with well-determined parameters and good Ti
EW measurements.  For example, from our observations of the solar spectrum, 
we derive $\log{\epsilon{\rm(Ti I)}} = 4.86$ and
$\log{\epsilon{\rm(Ti II)}} = 5.00$ before the adjustment, 
while \cite{ag89} give $\log{\epsilon{\rm (Ti)}} = 4.99$.

        The usually-measured Eu II lines were too weak to be measured in
the extremely alpha-weak star BD +80 245 (= HIP 40068).  As this star is very
interesting, and the r-process element Eu is a clue to
the history of this star, we used the Eu II lines given by \cite{s96}.
The Eu II lines and the measured EWs are:  EW(3819.67) = 2.4 m\AA, 
EW(3971.96) = 2.4 m\AA,
EW(4129.72) = 1.0 m\AA$\;$ (upper limit), and EW(4205.05) = 2.0 m\AA.  The
resulting abundance is [Eu/Fe] = $-0.76 \pm 0.25$.  Due to the extreme
weakness of the lines, it may be best to consider this abundance a
reliable upper limit.

        Figures 6(a)--(c) illustrate the frequency functions of this study for
 \teff, \logg{} and [Fe/H].   As can be seen, the survey is dominated by 
dwarfs with solar-like \teff{} values.   
The survey is split nearly evenly at [Fe/H] $= -1$, with 82 stars  
below and 86 stars above this value.  The mean [Fe/H] of the 168 
stars observed in this survey is $-1.21$.  
The minimum [Fe/H] value is $-3.01$ from 
HIP 50173 (= HD 88609), and the maximum [Fe/H] is $+0.02$ for the Sun.

\subsection{Estimating the Internal Errors}

        The internal uncertainties in the output abundances are due to
errors in the atmospheric parameters, atomic data (gf-values) and EW
measurements.  If we assume that the atmospheric parameters 
derived above are correct, we can estimate the uncertainty in the
final abundances due to the errors between gf-values and in the EW measurements:
\begin{equation}
\sigma_{rand}^2 = \langle\sigma(\log{\epsilon{\rm (X)}} )\rangle^2_{all\;stars}
\end{equation}
where $\sigma(\log{\epsilon{\rm (X)}})$ is the standard deviation of the mean 
of the abundance of element X given by the individual lines.  Note that we can
calculate this value only when we have multiple line measurements in a
given star.  For Li, where we only use the 6707 \AA$\;$ line, we have
adopted the $\sigma_{rand}$ of Fe I for this element.

        We use multiple observations of individual stars to quantify
the parameter-based uncertainties of our analysis methods.  The procedure is 
based on the assumption that the parameter-based error
can be expressed as a single value, and this value can
be calculated from a Monte Carlo-like approach.  Under this assumption,
the optimal way to determine these errors would be to take a large
number of spectra of the same star and analyse each spectra
using the same methods.  All the internal uncertainties  
would then be expressed in the variance of the stellar parameters and
abundances.

        Unfortunately, we do not have a large number of observations
of the same star, but we do have 18 stars with two or more spectra (41
spectra total).  These were used to estimate the internal uncertainties.
For each star with multiple observations, the variance between measurements
for each abundance was calculated by:

\begin{equation}
\sigma_{mult}^2 = \frac{\sum_{i=1}^{N} M_i(\sum_{j=1}^{M_i}\frac{\sqrt{(\overline{x}_i - x_j)^2}}{M_i})^2}{\sum_{i=1}^{N} M_i}
\end{equation}
where $N$ is the 18 stars, each with $M$ observations.  The value
$\overline{x}_i$ is the mean value of the abundance over the $M$ observations
of star $i$.  Note that we use this equation for both
the abundances (e.g. [X/H]) and the abundance ratios (e.g. [X/Fe]).
We also use Equation 2 to calculate the internal uncertainty of the
stellar parameters.  We find:  $\sigma(T_{eff}) = 40$ K,
$\sigma(\log g) = 0.06$, $\sigma$([Fe/H]$_{atm}$) $= 0.04$ and
$\sigma(\xi) = 0.11$ km/s.

        If we assume the errors $\sigma_{rand}$ and $\sigma_{mult}$ are
independent (in the case of the Fe lines, this is technically not true, 
as errors in the measurement of these lines can affect the choice of stellar 
parameters, but we assume that the effect is small), 
then we get the total internal error by adding in quadrature.  Table 7 lists 
these total errors as $\sigma$([X/H]) and $\sigma$([X/Fe]).

\subsection{Effects of Systematic Errors}

	As revealed in section 4.2, the stellar parameters derived here
show some systematic differences from the values derived by others.  These 
differences are probably inevitable, so it is important to understand the
effects of potental systematic errors in the parameters on the final derived
abundances.  As with the internal errors, knowing the potential amplitude of
the systematic errors places limits on the conclusions that can
be derived from the results.   

        To understand how systematic errors  affect the
derived abundance ratios, a series of tests were conducted on
a set of 13 stars.  These test stars were
selected to cover a wide range of temperatures, evolutionary states, and
metallicities.  For each test star, the same set of line measurements used 
in the abundance determinations were run through 10 different atmospheres, 
each with one or more parameters varied from the original values.  The first 8
were cases where only one parameter was varied:  $T_{eff} \pm 150$ K, 
\logg{} $\pm 0.2$ dex, [Fe/H]$_{atm} \pm 0.3$ dex, and $\xi \pm 0.3$ km/s.  
In the final two tests, the value of \teff{} was raised or
lowered 150 K and the other parameters were adjusted (following the method
described in section 4.1) to make the best fit possible given the
new \teff{}.  The mean net effect of the $\pm 150$ K \teff{} change on the other
parameters were:  \logg{} $\pm 0.3$ dex, [Fe/H]$_{atm} \pm 0.1$ dex, and
$\xi \pm 0.07$ km/s.  The change in $\xi$ varied widely from star to star,
while the changes in the other two parameters were more or less the same
star-to-star.

	For each star in a given test, the resulting abundances for each
species were compared to the abundances derived with the original parameters.
These differences were averaged over all 13 stars.  The results are listed 
in Table 8.  The columns labelled `All Vary' refer to the tests in 
which the \teff{} was varied by $\pm 150$ K and the other parameters were 
allowed to vary as well.  

	In section 4.1, we noted that the Kurucz 
atmospheres assume solar ratios for the metals, which does not accurately 
represent the abundance distribution of metal-poor stars.  To account for 
this, we adjusted the value of [Fe/H]$_{atm}$ upwards in order to increase the 
free electron supply.  As can be seen from Table 8, the net effect on the 
abundance ratios due to this change is fairly small.  The individual star 
with the largest changes due to adjusting [Fe/H]$_{atm}$ was HIP 57939 
(= HD 103095).  This star is moderately metal-poor ([Fe/H] $\sim -1.5$) and 
relatively cool (\teff{} = 4950 K) subdwarf.  The abundance ratio changes 
seen in this star were generally less than twice the listed values.  Note 
that one slightly sensitive species is Fe II, although the net effect would 
only require an adjustment in \logg{} of $\lesssim 0.1$ dex to restore the 
ionization equilibrium.

\section{Non-LTE Effects}

        \cite{ap99} compared \logg{} values for nearby stars derived via 
Hipparcos parallaxes to literature values obtained via spectroscopic analysis.  
Their procedure for calculating \logg{} was similar to what was used 
section 4.1, although the authors estimated the masses of the stars in 
their sample by their position on the CMD.  They find that for very 
metal-poor stars ([Fe/H] $< -2.0$) the value of the spectroscopic
\logg{} was significantly lower than the value obtained from the parallax.
The authors conclude that the assumption of LTE may not hold for these stars, 
as an ionizing UV photon can travel much farther in the stellar
atmosphere, leading over-ionization compared to LTE conditions.  

In Figure 7(a) we plot the difference between the \logg{} values 
from this work and \cite{ap99} against [Fe/H].  There are 3 metal-poor stars 
which have spectroscopic \logg{} values significantly lower than their 
trigometric \logg{} values.  To explore this further, in Figure 7(b) we plot 
the same quantities, except we expand the sample by using the Hipparcos-based 
\logg{} values calculated in this study (the ``\logg{} init.'' column in 
Table 5).  We exclude all stars that are suspected of binarity, have large 
values of the Hipparcos 
``goodness of fit'' parameter, have $\pi < 10$ mas, or have trigometric
$\sigma_{\logg{}} > 0.2$ dex (assuming the only source of error was from 
the measurement of the parallax).  The three low stars from Figure 6(b) 
are eliminated by these criteria (suspected of binarity), but one very 
metal-poor star (HIP 76976 = HD 140283) is added with a spectroscopic
\logg{} value $\sim 0.2$ dex lower than the Hipparcos-based \logg{} values.  
Overall, the spectroscopic \logg{} values are differ by 
$\Delta(log {\it g}) = -0.05 \pm 0.04$ (sdom, N = 71).

	If the claims of \cite{ap99} are correct, the 
\logg{} of dwarfs with [Fe/H] $< -2.0$ may need raising by several tenths of 
a dex.  Overall, this would lead to a systematic decrease of nearly all of the
abundance ratios.  Inspection of the 5 stars tested in section 5.2 with 
[Fe/H] $< -2.0$ shows the systematic changes in abundances are fairly
constant, independent of their temperature or evolutionary status.  For
example, for these 5 stars (HIP 14594, HIP 40068, HIP 68594, HIP 85855 and
HIP 87639) if \logg{} is increased 0.2 dex, the [Na/Fe] ratio changes by
$-0.09$, $-0.12$, $-0.07$, $-0.08$ and $-0.10$ dex, respectively.
These changes are somewhat significant when compared to more metal-rich 
stars, but the relative change between the metal-poor stars remains small.
       
     \cite{ti99} performed extensive detailed calculations
of the effect of departures from LTE in populating the energy levels of the
Fe I atom and concluded that non-LTE effects may plague abundances derived 
from Fe I lines, especially in hotter stars (\teff{} $> 6000$ K) and subdwarfs 
of low metallicity.  The increase in [Fe/H] over LTE determinations was as
large as 0.3 dex.  They also concluded that Fe abundances derived from Fe II 
lines, arising generally in atmospheric layers much deeper than those 
responsible for the Fe I lines, were in fact formed in LTE. 

     This last point is an important criterion in considering the reliability
of the abundances derived in this investigation. Our [Fe/H]-values are 
taken as a mean between $\log{\epsilon}({\rm Fe I})$ and 
$\log{\epsilon}({\rm Fe II})$, 
where the Fe II value is taken as fundamental, and the Fe I value is brought 
into agreement with it (within 0.03 dex) as a means of setting \logg. The 
value of \teff{} is set explicitly by means of the Fe I 
lines, via the excitation plot (as explained in Section 4). The question 
then is whether the [Fe/H]-values, based as they are on the assumption that 
\teff{} can be derived from LTE analysis of the Fe I lines, are reliable, i.e.,
satisfy these two criteria: (1) the abundance of Fe II should be fairly
independent of the exact value of \teff; (2) the spectroscopic \logg{} is in 
close agreement with the trigometric \logg{} for the 
choice of \teff{} derived by the Fe I excitation plot.

     We compare first with the \teff, \logg{} and [Fe/H] tabulations quoted
by \cite{ti99}, based on the catalog of \cite{t98}. There
are 35 stars in common, all dwarfs, with a range in metallicity from $-0.4$
to $-2.5$ in [Fe/H]. The differences, in the sense ``present minus 
\cite{ti99}'', are $\Delta T_{eff} = +15$ K $\pm\;20$ K (sdom), 
$\Delta\log{\it g} = -0.09 \pm 0.04$, and $\Delta$[Fe/H] $= -0.20 \pm 0.015$.
Confining attention to the eleven most metal-poor dwarfs does not change the 
result: $\Delta{T_{eff}} = +38$ K $\pm\;40$ K, $\Delta\log{\it g} = -0.06 \pm 0.05$ and $\Delta$[Fe/H] $= -0.19 \pm 0.03$. The 
\cite{ti99} \logg{} and [Fe/H] values are those ``adjusted'' for 
non-LTE effects in Fe I, and the latter turn out to be higher than those 
based on the conventional LTE results by about 0.2 dex.

     Except for [Fe/H], the changes are quite small. \cite{ti99} suggest that 
their values of \logg{} are close to those expected on the average based on the 
Hipparcos parallaxes. Our values of \logg{} are only slightly smaller, as
was seen in the comparison to Hipparcos-based \logg{} values above (Figure 6).
The difference is in the same direction and has a 
value (within the errors) much like that of our difference in \logg{} compared 
with \cite{ap99}.  

     The above argument assumes that our values of \teff{} are, of course,
accurate. The important point however, in the context of the present study, is
the effect of the choice of \teff{} on the abundance scale. The Fe II abundance
is fairly independent of \teff. We see that from inspection of Table 8,
where it is seen that the [Fe II/H]-abundance scarcely changes over a
300 K change in \teff. The reason for this has to do with the ionization
equilibrium of Fe: at optical depths greater than 0.1 in subdwarf atmospheres in
the \teff{} range from 5000--6200 K, the minority of the Fe is in the form of 
Fe I.  In deeper layers and in hotter stars, less than 10 percent of the 
Fe is in the form of Fe I.  Thus, changes of a few hundred degrees in \teff{} 
have little effect on $\log{\epsilon}({\rm Fe II})$.  Certainly the increase of 
38 K necessary to place us on the \teff{} scale of (say) \cite{a95} 
would have a negligible effect.

     Finally, we note the difference of 0.2 dex between our [Fe/H] values
and the non-LTE ``adjusted'' values of \cite{ti99}. The \cite{ti99} catalog 
contains [Fe/H]-values that are ``means'' of determinations from the 
literature, based on LTE analysis of (mostly) Fe I lines. They are subject, of 
course, to systematic errors in gf-values, EWs, effective temperatures, etc, 
and we have no way 
to judge those.  What is valuable in \cite{ti99} is not the assumed set of 
``absolute'' LTE [Fe/H] values, but rather the corrections to those values 
based on non-LTE considerations.

\section{Summary}

        We have observed and analyzed elemental abundances for 168 stars 
(mostly dwarfs) using 191 high-resolution, high-S/N
spectra.  The methods used to determine the parameters and abundances
are self-consistent and should provide accurate relative abundances
for the target stars.  The overall precision of the abundances
ultimately depends on how well the basic assumptions of the analysis hold,
which are: 1) the abundances of the elements considered can be determined
accurately by plane-parallel atmospheres using an LTE analysis,  2) the
\teff{} of the stars can be determined by the plot of abundance vs
excitation potential for Fe I, 3) the surface gravity can be determined by
matching iron abundances determined by the neutral and singly ionized
lines, 4) the microturbulent velocity can be determined by the abundance
vs line strength plot for Fe I, and 5) the van der Waals damping parameter
can be accurately determined by the \cite{u55} approximation.

	To test these assumptions, we have compared the stellar parameters 
we derive against independent measurements, and we study the effects of 
parameter variations on the resulting abundances.  Our \teff{} scale is 
consistent with independent determinations (including the non-LTE scale of
Thevenin \& Idiart 1999) to within a small offset.  Our surface gravities 
also compare well with Hipparcos-based values, although we lack data for
stars with [Fe/H] $< -2$, which is where \cite{ap99} suggest non-LTE effects
exist.  More importantly, we show that any reasonable variation in the
derived parameters does not cause drastic changes in the derived abundance
ratios.  

\acknowledgments

The papers of this series make up the PhD thesis of JPF.  
JPF is grateful to R. P. Kraft, R. Peterson, M. Bolte, and P. Guhathakura
for their insights, advice and efforts on his PhD committee.  He also wishes
to acknowledge M. Shetrone, J. Johnson, and T. Misch for their thoughtful
comments and assistance with observations.   Special thanks goes to the
anonymous referee who provided several useful comments.
This research was supported by NSF Contract AST 96-18351 to RPK. 


\clearpage 

\newpage
\figcaption[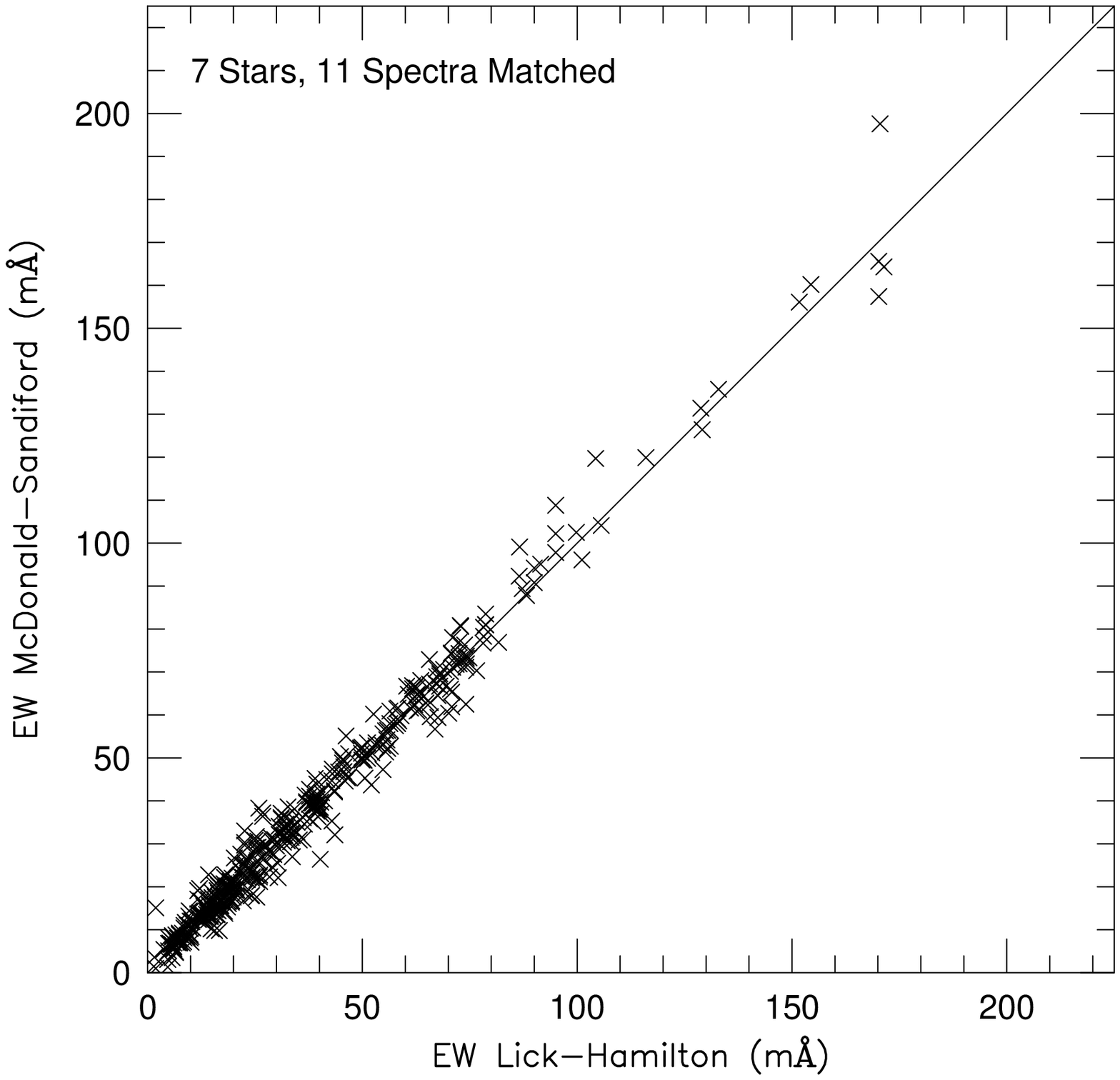]{Comparison of the equivalent width (EW) 
measurements between the Sandiford spectrograph (McDonald 2.1-m) and Hamilton 
spectrograph (Lick 3-m and CAT).\label{f1}}  

\figcaption[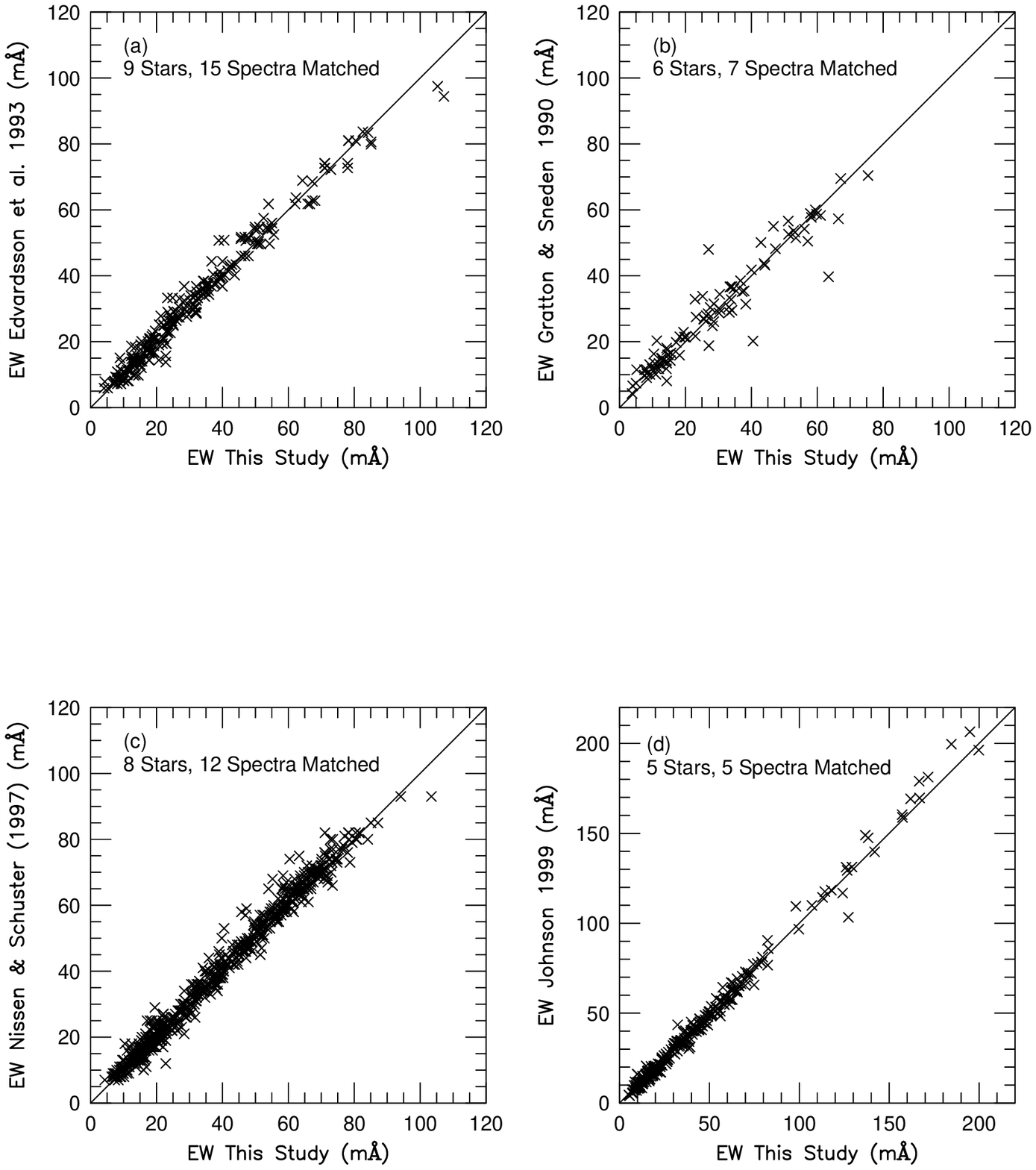]{Comparisons of equivalent width (EW) 
measurements from this paper and (a) \cite{ed}, (b) \cite{gs90}, 
(c) \cite{ns97} and (d) \cite{jj99}.\label{f2}} 

\figcaption[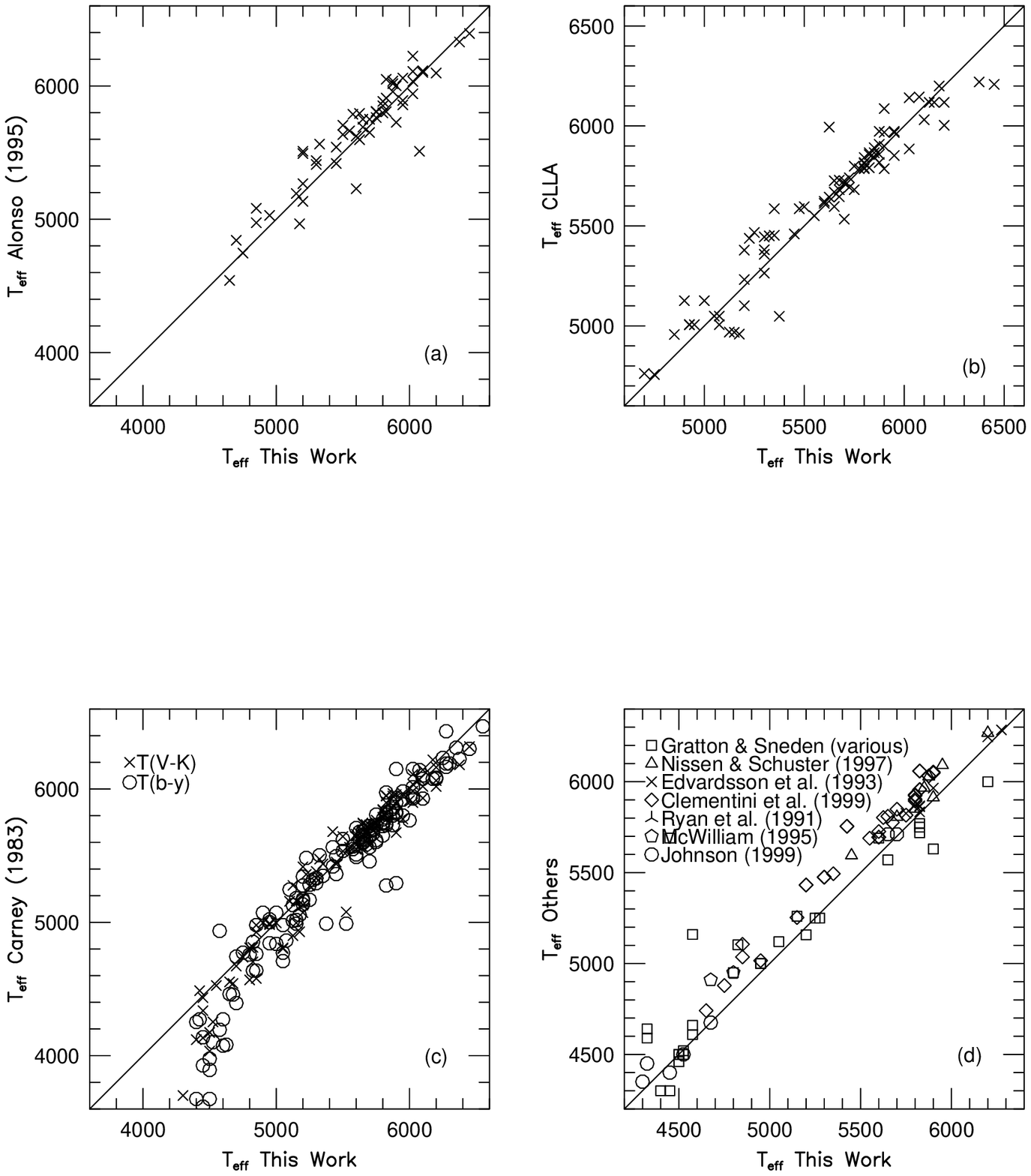]{\teff{} value comparisons between this work 
and (a) \cite{a95}, (b) \cite{clla}, (c) \cite{c83} and (d) various literature 
studies.\label{f3}}

\figcaption[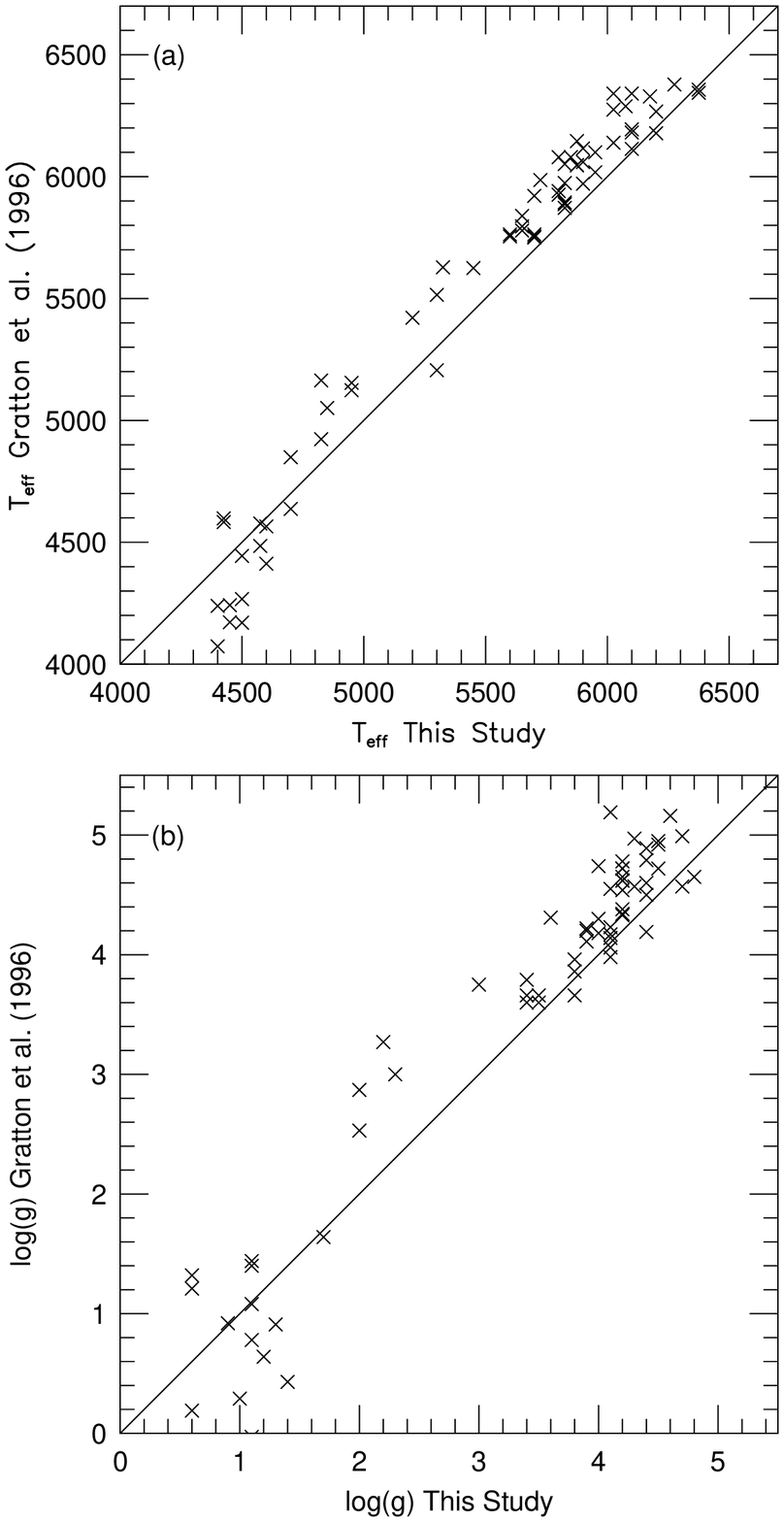]{(a) \teff{} and (b) \logg{} value comparisons 
between this work and \cite{g96}.  The structure seen at low temperatures
and surface gravities (also seen in Figure 3(c) seem to be due to
differences between the photometric and spectroscopic \teff{} scales for 
giant stars. \label{f4}}  

\figcaption[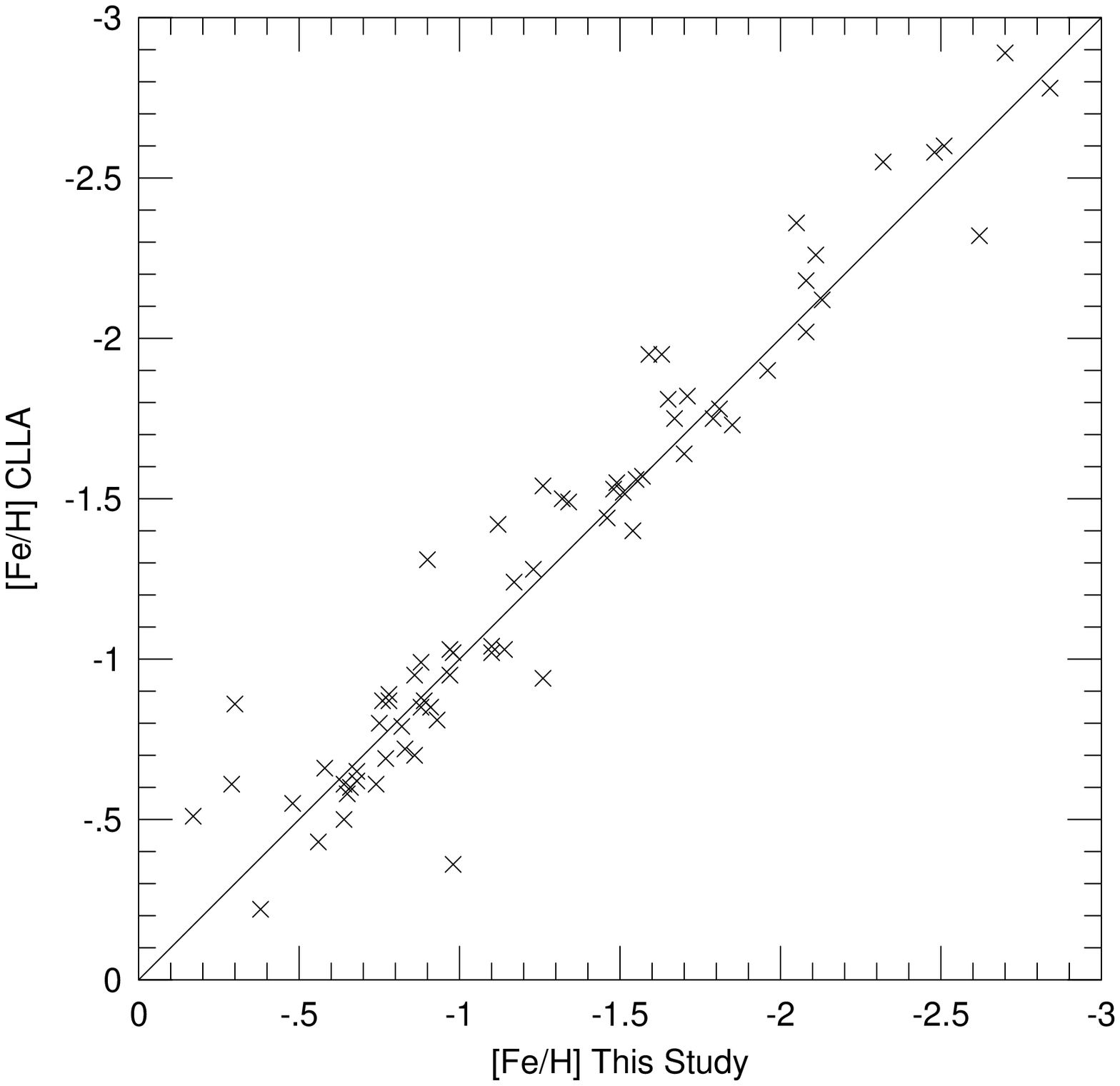]{Comparison of the value of [Fe/H] derived by 
the \cite{clla} survey and this work.\label{f5}}

\figcaption[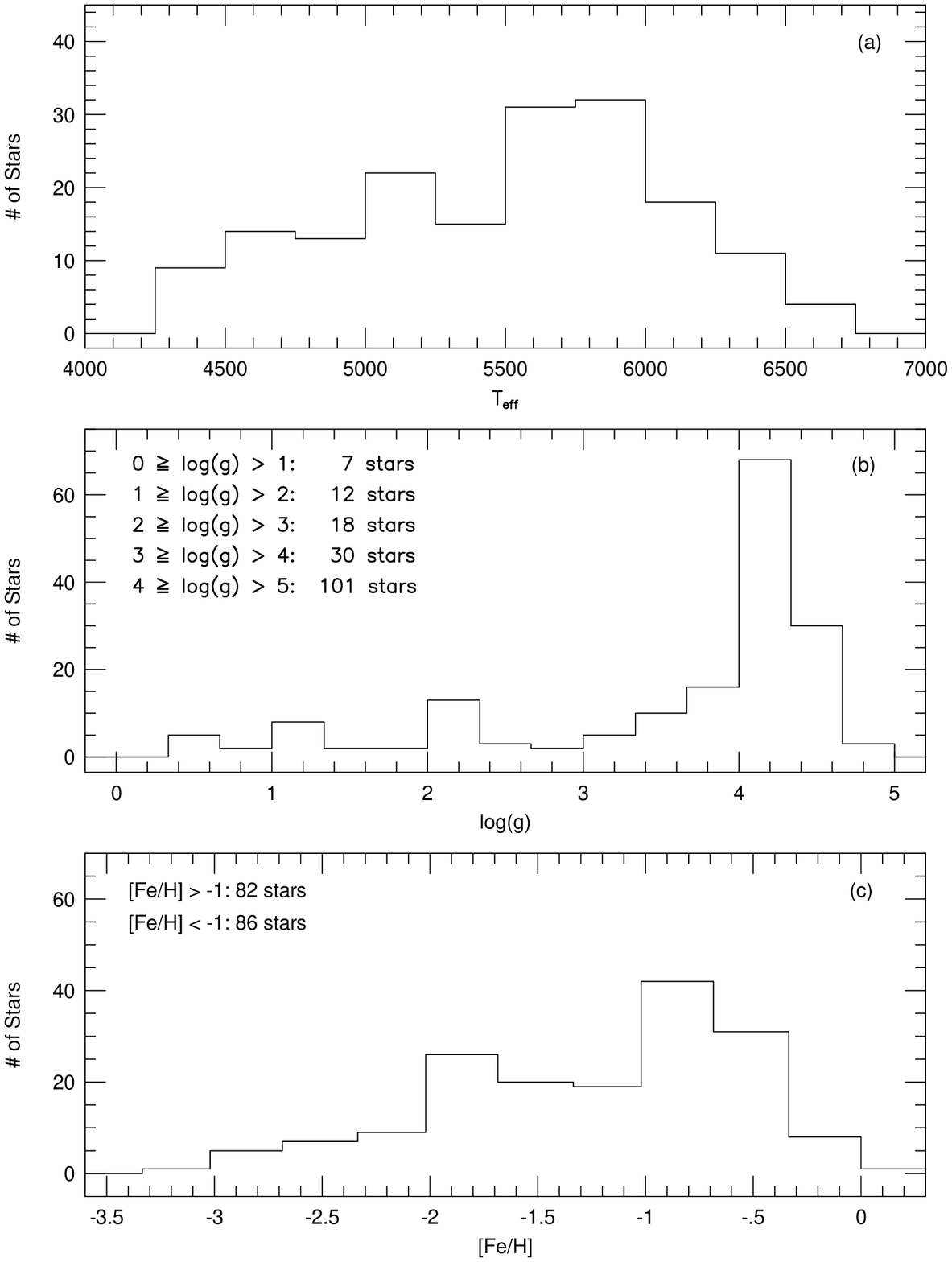]{Histograms of the various stellar parameters.  
(a) and (b) show that the most common star in the survey is a solar-like 
metal-deficient dwarf, although there a number of subgiants and giants 
included in the survey as well. \label{f7}  }

\figcaption[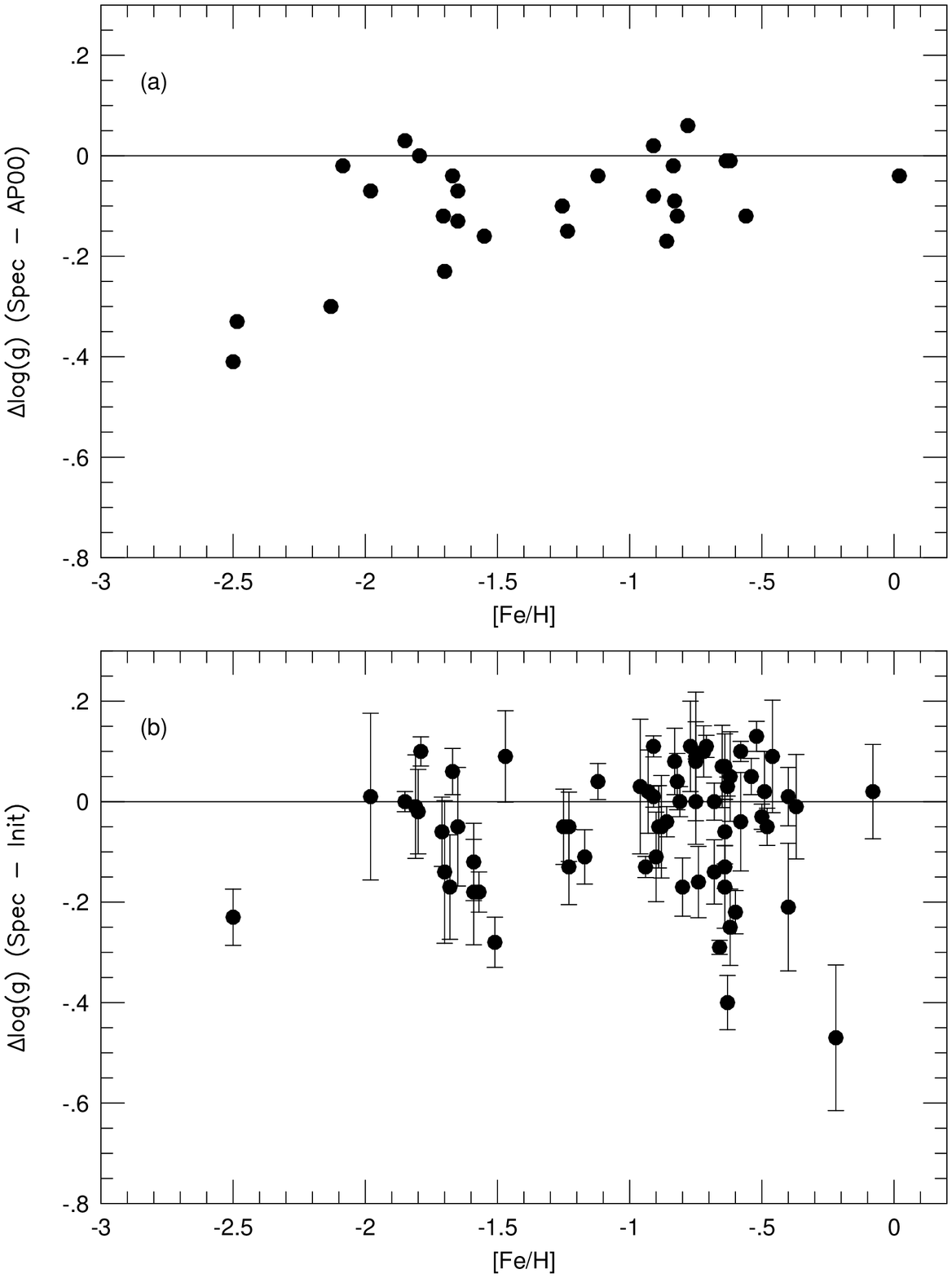]{Comparisons between the value of \logg{} 
derived spectroscopically in this work against the value derived from Hipparcos 
parallax distances (the `trigometric' \logg).  Panel (a) compares against the 
values determines by \cite{ap99}, while panel (b) compares the spectroscopic 
result of this work against the initial trigometric estimate, for those stars 
listed in Table 6 with reliable parallax values.\label{f6}}

\plotone{Fulbright.fig1.ps}
 
\plotone{Fulbright.fig2.ps}
 
\plotone{Fulbright.fig3.ps}
 
\plotone{Fulbright.fig4.ps}
 
\plotone{Fulbright.fig5.ps}
 
\plotone{Fulbright.fig6.ps}
 
\plotone{Fulbright.fig7.ps}
 
\begin{deluxetable}{rrrrl}
\tablenum{1}
\normalsize
\tablewidth{0pt}
\tablecaption{Log of Observing Runs}
\tablehead{
\colhead{Run}&\colhead{Dates}&\colhead{Telescope}&\colhead{Instrument}&\colhead{$\lambda\lambda$ Coverage (approx.)} \\}
\startdata
Sep94 & 18 September 1994 & Lick 3-m & Hamilton & 4460--7090 \AA$\;$ (incomplete) \nl
Nov94 & 12 November 1994 & Lick 3-m & Hamilton & 4460--7090 \AA$\;$ (incomplete) \nl
Jun95 & 10 June 1995 & Lick 3-m & Hamilton & 3900--10075 \AA$\;$ \nl
Oct97 & 18--20 October 1997 & Lick 0.6-m CAT & Hamilton & 3650--10750 \AA$\;$ \nl
ESO   & 24 November 1997 & ESO 3.6-m & CASPEC & 5700--8160 \AA$\;$ (incomplete) \nl
Mar98 & 17--18 March 1998 & Lick 3-m & Hamilton & 3700--10075 \AA$\;$ \nl
K98 & 26 March 1999 & Keck I 10-m & HIRES & 4380--6780 \AA$\;$ (incomplete)\nl
Jun98 & 7--8 June 1998 & Lick 3-m & Hamilton & 3700--10075 \AA$\;$ \nl
Jul98 & 28 July 1998 & Lick 3-m & Hamilton & 3800--8850 \AA$\;$ \nl
Aug98 & 8--9 August 1998 & Lick 3-m & Hamilton & 4140--9900 \AA$\;$ \nl
A98 & 10--11 August 1998 & Lick 3-m & Hamilton & 3800--8850 \AA$\;$ \nl
Oct98 & 30--31 October 1998 & Lick 3-m & Hamilton & 4140--9900 \AA$\;$ \nl
Jan99 & 26--31 January 1999 & McDonald 2.1-m & Sandiford & 5300--6500 \AA$\;$ \nl
K99 & 23 March 1999 & Keck I 10-m & HIRES & 3760--6180 \AA$\;$ (incomplete)\nl
Apr99 & 6--7 April 1999 & Lick 3-m & Hamilton & 4140--9900 \AA$\;$ \nl
May99 & 24 May 1999 & Lick 3-m & Hamilton & 4140--9900 \AA$\;$ \nl
\enddata
\end{deluxetable}
\clearpage
\begin{deluxetable}{lccc}
\tablenum{7}
\tablewidth{0pt}
\tablecaption{Random Uncertainties in Abundances}
\tablehead{
\colhead{Species} & \colhead{$\sigma_{rand}$} & \colhead{$\sigma_{\rm [X/H]}$} & \colhead{$\sigma_{\rm [X/Fe]}$} }
\startdata
Fe I  & 0.063   &   0.074     &        \nl
Fe II & 0.074   &   0.086     &        \nl
$\langle$Fe$\rangle$ &         &   0.079     &        \nl
Li I  &        &    0.126  & 0.100  \nl
Na I  & 0.063   &   0.100     & 0.103  \nl
Mg I  & 0.087   &   0.092     & 0.070  \nl
Al I  & 0.087   &   0.092     & 0.094  \nl
Si I  & 0.059   &   0.064     & 0.072  \nl
Ca I  & 0.063   &   0.077     & 0.073  \nl
Ti I  & 0.055   &   0.080     &             \nl  
Ti II & 0.056   &   0.085     &        \nl
$\langle$Ti$\rangle$  &         &             & 0.087  \nl
V  I  & 0.122   &   0.136     & 0.129  \nl
Cr I  & 0.057   &   0.075     & 0.075  \nl
Ni I  & 0.062   &   0.070     & 0.068  \nl
Y  II & 0.059   &   0.072     & 0.064  \nl
Zr II & 0.098   &   0.118     & 0.113  \nl
Ba II & 0.086   &   0.119     & 0.117  \nl
Eu II & 0.096   &   0.100     & 0.105  \nl
        &    &   & $\sigma_{\rm ratio}$  \nl
[Na/Mg] &    &   &     0.119  \nl
[Y/Zr]  &    &   &     0.125  \nl
[Ba/Eu] &    &   &     0.140  \nl
[$\alpha$/Fe]  &    &   &     0.119  \nl 
\enddata
\end{deluxetable}%

\end{document}